\documentclass[a4paper]{article}
\usepackage[leftcaption]{sidecap}
\usepackage{url}
\usepackage{graphicx}
\usepackage{mathptmx}      
\usepackage{latexsym}
\usepackage{amssymb,amsmath,bm,color,threeparttable,rotating,dcolumn,bbding, pifont, fancyhdr,bm}
\usepackage[colorlinks=true,linkcolor=blue,citecolor=blue,filecolor=blue]{hyperref}
\usepackage[round]{natbib}\usepackage{amssymb,amsmath,bm,color}
\usepackage{booktabs,dcolumn}
\usepackage{float}
\usepackage[affil-it]{authblk}
\newcommand{\BE}{\begin{equation}}
\newcommand{\EE}{\end{equation}}
\newcommand{\tr}{\mbox{tr}}

\newcommand{\Lower}[1]{\smash{\lower 1ex \hbox{#1}}}

\def\Aset{\mathbb{A}}

\begin{document}
\title{Robust classification via finite mixtures of  matrix-variate skew $t$ distributions}
\author{Abbas Mahdavi \thanks{Email: \texttt{a.mahdavi@vru.ac.ir}}}
\affil{Department of Statistics, Vali-e-Asr University of Rafsanjan, Rafsanjan, Iran.}

\author{Narayanaswamy Balakrishnan\thanks{Email: \texttt{bala@mcmaster.ca}}}
\affil{Department of Mathematics and Statistics, McMaster University, Hamilton, ON, Canada.}

\author{Ahad Jamalizadeh \thanks{Email: \texttt{a.jamalizadeh@uk.ac.ir}}} 
\affil{
Department of Statistics, Faculty of
Mathematics \& Computer, Shahid Bahonar University of Kerman, Kerman, Iran.}

\date{}
\maketitle
\begin{abstract}
Analysis of matrix-variate data is becoming increasingly common in the literature, particularly in the field of clustering and classification. It is well-known that real data, including real matrix-variate data, often exhibit high levels of asymmetry. To address this issue, one common approach is to introduce a tail or skewness parameter to a symmetric distribution. In this regard, we introduced here a new distribution called the matrix-variate skew $t$ distribution (MVST), which provides flexibility in terms of heavy tail and skewness. 
We then conduct a thorough investigation of various characterizations and probabilistic properties of the MVST distribution. We also explore extensions of this distribution to a finite mixture model. To estimate the parameters of the MVST distribution, we develop an efficient EM-type algorithm that computes maximum likelihood (ML) estimates of the model parameters. 
To validate the effectiveness and usefulness of the developed models and associated methods, we perform empirical experiments using simulated data as well as three real data examples. Our results demonstrate the efficacy of the developed approach in handling asymmetric matrix-variate data.
\end{abstract}

\section{Introduction }
The advent of modern data collection technologies, such as electronic sensors, cell phones and web browsers, has resulted in an abundance of multivariate data sources. Much of these data can be represented as a matrix-variate (three-way) data with two ways associated to the row and column dimension of each
matrix-variate observation and the third one representing subjects (see \cite{kroonenberg2008applied}). Matrix data can occur in different application domains such as spatial multivariate data, longitudinal data on multiple response variables, or spatio-temporal data. For this reason, statistical methods that can effectively utilize three-way data have become increasingly popular. The matrix-variate normal (MVN) distribution is one of the most commonly used matrix-variate elliptical distributions. However, for many real phenomena, the tails of the MVN distribution are lighter than required, with a direct impact on the corresponding mixture model. In particular, in robust statistical analysis, heavy-tailed distributions are essential, and these include slash and $t$ distributions. Matrix-variate $t$ (MVT) distribution has been discussed by \cite{dickey1967matricvariate} and some distributional properties of it has also been studied by \cite{dickey1967matricvariate}.

Flexibility and robustness are often lacking in symmetric models when dealing with highly asymmetric data. To address this issue, a recognized method is to add a tail or skewness parameter to a symmetric distribution. Several formulations have been discussed in the literature in the form of continuous mixtures of normal variables, where a mixing variable operates on the mean or on the variance, or on both mean and  variance of a multivariate normal variable. \cite{arellano2021formulation} presented a general formulation which encompasses a large number of existing constructions involving continuous mixtures of normal variables. Given a real-valued function $r(u,w)$ and a positive-valued function $s(u,w)$, a generalized mixture of a $p$ variate normal distribution is given by
\begin{align}\label{general}
\bm Y\overset{d}{=}\bm\xi+r(U,W)\bm\gamma+s(U,W)\bm X,
\end{align}
where $\overset{d}{=}$ denotes equality in distribution, $\bm\xi\in\mathbb{R}^p$, $\bm\gamma\in\mathbb{R}^p$, $\bm X\sim N_p(\bm 0,\bm\Sigma)$ and $U$ and $W$ are univariate random variables, with $(\bm X,U,W)$ being mutually independent.

The representation in \eqref{general} can be extended to the matrix-variate case as
\begin{align}\label{general_matrix}
\bm Y\overset{d}{=}\bm M+r(U,W)\bm\Lambda+s(U,W)\bm Z,
\end{align}
where $\bm M$ and $\bm\Lambda$ are $n\times p$ matrices representing the location and
skewness, respectively, $\bm Z\sim N_{n\times p}(\bm 0,\bm\Sigma,\bm\Psi)$ and $(\bm Z,U,W)$ are mutually independent. It is worth noting that  the univariate nature of functions $r(u,w)$ and $s(u,W)$ simplifies the stochastic representation in \eqref{general_matrix},  leading to more suitable properties for $\bm Y$ and also facilitates easier parameter estimation. Furthermore, the representation in \eqref{general} can be considered, after rearranging into a vector (denoted by $\mbox{Vec}(\bm Y)$), as
\begin{align}
\mbox{Vec}(\bm Y)\overset{d}{=}\mbox{Vec}(\bm M)+r(U,W)\mbox{Vec}(\bm\Lambda)+s(U,W)\mbox{Vec}(\bm Z).
\end{align}

However, to address highly asymmetric data and to have more flexibility, the functions $r(u,w)$ and $s(u,w)$ can also be considered as $p$-dimensional random variables. This approach may introduce complexity in both density function and parameter estimation issues. For instance, \cite{rezaei2020scale} extended the scale and shape mixtures of multivariate skew normal distributions to a matrix-variate setting, and studied special cases and their properties. Here, we concentrate on simplifying the proposed model and the associated estimation procedure by focusing on the univariate case of $r(u,w)$ and $s(u,w)$. However, it should be noted that the proposed model is not a special case of \cite{rezaei2020scale}.

Based on \eqref{general_matrix}, various cases have been introduced. For example, \cite{gallaugher2017matrix} introduced a matrix-variate skew-$t$ distribution using the following matrix-variate normal variance-mean mixture representation:
\begin{equation}\label{var-mean}
\bm Y\overset{d}{=}\bm M+W\bm\Lambda+W^{1/2}\bm Z,
\end{equation}
where $\bm M$ and $\bm\Lambda$ are $n\times p$ matrices representing the location and
skewness, respectively, $\bm Z\sim N_{n\times p}(\bm 0,\bm\Sigma,\bm\Psi)$, $W\sim IG(\nu/2,\nu/2)$ and $IG(\cdot)$ denotes the inverse gamma distribution. The resulting density function of $\bm Y$ in \eqref{var-mean} is given by
\begin{align}
&f(\bm Y;\bm M,\bm\Sigma,\bm\Psi,\bm\Lambda,\nu)\notag\\&=\frac{2(\nu/2)^{\nu/2}\exp\big\{\tr[\bm\Sigma^{-1}(\bm Y-\bm M)\bm\Psi^{-1}\bm\Lambda^\top]}{(2\pi)^{np/2}\vert\bm\Sigma\vert^{p/2}\vert\bm\Psi\vert^{n/2}\Gamma(\nu/2)}\bigg(\frac{\delta(\bm Y;\bm M,\bm\Sigma,\bm\Psi)+\nu}{\rho(\bm\Sigma,\bm\Psi,\bm\Lambda)}\bigg)^{-\frac{\nu+np}{4}}\notag\\&\times\kappa_{-(\nu+np)/2}\bigg(\sqrt{\rho(\bm\Sigma,\bm\Psi,\bm\Lambda)(\delta(\bm Y;\bm M,\bm\Sigma,\bm\Psi)+\nu)}\bigg),\ \ \ \bm Y\in\mathbb{R}^{n\times p},
\end{align}
where 
\begin{align*}
\delta(\bm Y;\bm M,\bm\Sigma,\bm\Psi)&=\tr[\bm\Sigma^{-1}(\bm Y-\bm M)\bm\Psi^{-1}(\bm Y-\bm M)^\top],\\ 
\rho(\bm\Sigma,\bm\Psi,\bm\Lambda)&=\tr[\bm\Sigma^{-1}\bm\Lambda\bm\Psi^{-1}\bm\Lambda^\top],
\end{align*}
and $\kappa_x$ is the modified Bessel function of the third kind with index $x$.
From \eqref{var-mean}, some other matrix-variate skew distributions can be introduced by assuming different distributions for $W$. For more details, one may refer to \cite{gallaugher2018finite}.

\cite{naderi2020theoretical} introduced a new family of matrix-variate distributions based on the matrix-variate mean-mixture of normal (MVMMN) distributions, as 
\begin{align}\label{Naderi}
\bm Y\overset{d}{=}\bm M+W\bm\Lambda+\bm Z.
\end{align}
\sloppy{Based on \eqref{Naderi}, three special cases, including the restricted matrix-variate skew-normal (RMVSN), exponentiated MVMMN (MVMMNE) and  mixed-Weibull MVMMN   (MVMMNW) distributions, have been studied by using half normal, exponential and Weibull distributions for $W$, respectively.} Several other skew matrix-variate distributions have also been  discussed in the literature; see \cite{chen2005matrix},~\cite{dominguez2007matrix} and \cite{zhang2020graphical}.

\sloppy{One common statistical challenge faced by researchers is identifying sub-populations or clusters within multivariate data.} Recently, researchers have explored the use of finite mixture models for matrix-variate data in applications such as image analysis, genetics, and neuroscience. These models offer a flexible framework for capturing complex patterns in the data and can provide insights into the underlying sub-populations or clusters; see  \cite{viroli2011finite}, \cite{gallaugher2018finite},   \cite{thompson2020classification} and   \cite{tomarchio2020two}.

In this work, we introduce and study in detail finite mixtures of a simple matrix-variate skew-$t$ (FM-MVST) distribution, based on \eqref{general_matrix}, for dealing with clustering and classification of asymmetric and heavy-tailed matrix-variate data. The proposed model's simplicity in both the density function and stochastic representation leads to a convenient strategy for parameter estimation using the expectation conditional maximization either (ECME; \cite{liu1994ecme}) algorithm, which is a variant of the EM algorithm \citep{dempster1977maximum}.
In addition, using simulated and real datasets, we show how the proposed EM algorithm can be implemented for determining the ML estimates of the model parameters for the finite mixture of the proposed model.

The rest of this paper is organized as follows. Section~\ref{sec:2} presents the formulation of  MVST distribution and discusses how the ECME algorithm can be proposed for the ML estimation of model parameters. In Section~\ref{sec-FM}, the finite mixture of MVST (FM-MVST) distributions is defined and then the implementation of the EM algorithm for fitting the FM-MVST model is presented. The proposed methods are illustrated by two simulation studies in Section~\ref{sec_sim} and also by the analysis of three real data sets in Section \ref{sec_data}. Finally, some concluding remarks and possible avenues for future work are outlined in
Section~\ref{sec:con}.
\section{Methodology}\label{sec:2}
\subsection{The model}
A $n\times p$-variate random matrix $\bm Y$ is said to have a matrix-variate skew-$t$ (MVST) distribution, with $n\times p$ location matrix $\bm M$, $n\times n$ and $p\times p$ scale matrices $\bm\Sigma$ and $\bm\Psi$, $n\times p$ shape matrix $\bm\Lambda$ and flatness parameters $\nu$, if its probability density function (pdf) is
\begin{align}\label{MVST_pdf}
&f_{MVST}(\bm Y;\bm\theta)\notag\\&=\frac{2(\nu/2)^{\nu/2}\Gamma\big(\frac{\nu+np}{2}\big)\vert\bm\Sigma\vert^{-p/2}\vert\bm\Psi\vert^{-n/2}}{(2\pi)^{np/2}\Gamma(\nu/2)\sqrt{\rho(\bm\Sigma,\bm\Psi,\bm\Lambda)+1}}\bigg(\frac{\delta(\bm Y;\bm M,\bm\Sigma,\bm\Psi)+\nu-\Delta^2(\bm Y;\bm\theta)}{2}\bigg)^{-\frac{\nu+np}{2}}\notag\\&\times T_{(\nu+np)}\bigg(\Delta(\bm Y;\bm\theta)\sqrt{\frac{\nu+np}{\delta(\bm Y;\bm M,\bm\Sigma,\bm\Psi)+\nu-\Delta^2(\bm Y;\bm\theta)}}~\bigg),\ \ \ \bm Y\in\mathbb{R}^{n\times p},
\end{align}
where $\bm\theta=(\bm M,\bm\Sigma,\bm\Psi,\bm\Lambda,\nu)$ denots all the model parameters, $\Delta(\bm Y;\bm\theta)=\tr[\bm\Sigma^{-1}(\bm Y-\bm M)\bm\Psi^{-1}\bm\Lambda^\top]/\sqrt{\rho(\bm\Sigma,\bm\Psi,\bm\Lambda)+1}$ and $T_\nu(\cdot)$ denotes the cumulative distribution function (cdf) of Student's $t$ distribution with $\nu$ degrees of freedom. The MVST distribution reduces to the RMVSN distribution \citep{naderi2020theoretical} when $\nu\rightarrow\infty$.

Moreover, the MVST distribution possess the stochastic representation
\begin{eqnarray}\label{con_rep}
\bm Y\overset{d}{=}\bm M+W^{-1/2}\big(U\bm\Lambda+\bm Z\big),
\end{eqnarray}
where $\bm Z\sim N_{n\times p}(\bm 0,\bm\Sigma,\bm\Psi)$, $W\sim\Gamma(\nu/2,\nu/2)$ and
$U\sim TN(0,1)I_{(0,\infty)}$. Herein, $TN(\mu, \sigma^2)I_{\Aset}$ represents a doubly truncated normal distribution defined in the interval $\Aset=\{a_1 < x < a_2\}$, and $I_{\Aset}$ denotes the indicator function of set $\Aset$. Form \eqref{con_rep}, it is easy to show that
\begin{align*}
E(\bm Y)&=\bm M+\frac{\Gamma\big(\frac{\nu-1}{2}\big)}{\Gamma\big(\frac{\nu}{2}\big)}\bigg(\frac{\nu}{\sqrt{\pi}}\bigg)\bm\Lambda,\\
\mbox{Vec}(\bm Y)&\sim rST_{np}\big(\mbox{Vec}(\bm M),\bm\Psi\otimes\bm\Sigma,\mbox{Vec}(\bm\Lambda)\big),
\end{align*}
where $\otimes$ is the Kronecker product and 
 $rST_p$ denotes the $p$-variate restricted skew-$t$ distribution (see; \cite{lin2015robust} and \cite{lee2016finite}). 

The stochastic representation given in \eqref{con_rep} not only facilitates random number generation, but also enables the implementation of EM algorithm for determining the maximum likelihood (ML) estimates of the parameters of the MVST distribution. This leads to the hierarchical representation
\begin{align}\label{hier_rep}
\bm Y\vert(\gamma,w)&\sim N_{n\times p}(\bm M+\gamma\bm\Lambda,w^{-1}\bm\Sigma,\bm\Psi),\notag\\
\gamma\vert w&\sim TN(0,w^{-1})I_{(0,\infty)},\notag\\
W&\sim\Gamma(\nu/2,\nu/2),
\end{align}
where $\gamma=W^{-1/2}U$ and $W$ are treated as latent variables.
Then, $(\bm{Y}^\top,W,\bm\gamma)^\top$ has the  joint pdf as
\begin{align}\label{joint_pdf}
f_{\bm Y,W,\gamma}(\bm Y,w,\gamma)=2w^{1/2}\phi_{n\times p}(\bm Y;\bm M+\gamma\bm\Lambda,w^{-1}\bm\Sigma,\bm\Psi)~\phi(w^{1/2}\gamma)~g(w;\nu/2,\nu/2),
\end{align}
where $\phi(\cdot)$ and $\phi_{n\times p}(\cdot;\bm M,\bm\Sigma,\bm\Psi)$ are the pdfs of $N(0,1)$ and $N_{n\times p}(\bm M,\bm\Sigma,\bm\Psi)$, respectively, and $g(\cdot,\alpha,\beta)$ denotes the pdf of the
gamma distribution with mean $\alpha/\beta$.

Integrating out $W$ and $\gamma$, respectively, from \eqref{joint_pdf}, we obtain the joint pdfs
\begin{align}\label{joint_ygamma}
f_{\bm Y,\gamma}(\bm Y,\gamma)&=\frac{2(\nu/2)^{\nu/2}\Gamma\big(\frac{\nu+np+1}{2}\big)\vert\bm\Sigma\vert^{-p/2}\vert\bm\Psi\vert^{-n/2}}{(2\pi)^{(np+1)/2}\Gamma(\nu/2)}\notag\\&\bigg(\frac{\delta(\bm Y;\bm M,\bm\Sigma,\bm\Psi)+\big(\rho(\bm\Sigma,\bm\Psi,\bm\Lambda)+1\big)\gamma^2-2\eta(\bm Y;\bm\theta)\gamma+\nu}{2}\bigg)^{-\frac{\nu+np+1}{2}},
\end{align}
where $\eta(\bm Y;\bm\theta)=\tr[\bm\Sigma^{-1}(\bm Y-\bm M)\bm\Psi^{-1}\bm\Lambda^\top]$,
and
\begin{align}\label{joint_yw}
f_{\bm Y,W}(\bm Y,w)&=\frac{2(\nu/2)^{\nu/2}\vert\bm\Sigma\vert^{-p/2}\vert\bm\Psi\vert^{-n/2}}{(2\pi)^{np/2}\Gamma(\nu/2)\sqrt{\rho(\bm\Sigma,\bm\Psi,\bm\Lambda)+1}}w^{\frac{\nu+np}{2}-1}\notag\\&\times\exp\bigg\{-\frac{w\big(\delta(\bm Y;\bm M,\bm\Sigma,\bm\Psi)+\nu-\Delta^2(\bm Y;\bm\theta)\big)}{2}\bigg\}\Phi\big(w^{1/2}\Delta(\bm Y;\bm\theta)\big),
\end{align}
where $\Phi(\cdot)$ denotes the cdf of the standard normal distribution.

Dividing \eqref{joint_pdf} by \eqref{joint_ygamma}, we obtain
\begin{align}
&W\vert(\bm Y,\gamma)\notag\\&\sim\Gamma\bigg(\frac{\nu+np+1}{2},\frac{\delta(\bm Y;\bm M,\bm\Sigma,\bm\Psi)+\big(\rho(\bm\Sigma,\bm\Psi,\bm\Lambda)+1\big)\gamma^2-2\eta(\bm Y;\bm\theta)\gamma+\nu}{2}\bigg).
\end{align}
Additionally, dividing \eqref{joint_pdf} by \eqref{joint_yw}, we get
\begin{align}
\gamma\vert(\bm Y,w)\sim TN\bigg(\frac{\eta(\bm Y;\bm\theta)}{\rho(\bm\Sigma,\bm\Psi,\bm\Lambda)+1},\frac{w^{-1}}{\rho(\bm\Sigma,\bm\Psi,\bm\Lambda)+1}\bigg)I_{(0,\infty)}.
\end{align}
From \eqref{MVST_pdf} and \eqref{joint_yw}, it is easy to see that
\begin{align}\label{w.y_pdf}
f(w\vert\bm Y)=C~w^{\frac{\nu+np}{2}-1}\exp\bigg\{-\frac{w\big(\delta(\bm Y;\bm M,\bm\Sigma,\bm\Psi)+\nu-\Delta^2(\bm Y;\bm\theta)\big)}{2}\bigg\}\Phi\big(w^{1/2}\Delta(\bm Y;\bm\theta)\big),
\end{align}
where 
\begin{equation}
C=\frac{\bigg(\frac{\delta(\bm Y;\bm M,\bm\Sigma,\bm\Psi)+\nu-\Delta^2(\bm Y;\bm\theta)}{2}\bigg)^{\frac{\nu+np}{2}}}{\Gamma\big(\frac{\nu+np}{2}\big)T_{(\nu+np)}\big(\Delta(\bm Y;\bm\theta)\sqrt{\frac{\nu+np}{\delta(\bm Y;\bm M,\bm\Sigma,\bm\Psi)+\nu-\Delta^2(\bm Y;\bm\theta)}}~\big)}.
\end{equation}
Further, by using \eqref{MVST_pdf} and \eqref{joint_ygamma}, it can be shown that
\begin{align}
\gamma\vert\bm Y\sim Tt\bigg(\frac{\eta(\bm Y;\bm\theta)}{\rho(\bm\Sigma,\bm\Psi,\bm\Lambda)+1},\frac{\delta(\bm Y;\bm M,\bm\Sigma,\bm\Psi)+\nu-\Delta^2(\bm Y;\bm\theta)}{(\rho(\bm\Sigma,\bm\Psi,\bm\Lambda)+1)(\nu+np)},\nu+np\bigg)I_{(0,\infty)},
\end{align}
where $Tt(\mu, \sigma^2,\nu)I_{\Aset}$ represents a doubly truncated $t$ distribution with $\nu$ degrees of freedom defined in the interval $\Aset=\{a_1 < x < a_2\}$.
From the conditional density in \eqref{w.y_pdf}, we find
\begin{align}
E(W\vert \bm Y)=C_0\frac{T_{(\nu+np+2)}\big(\Delta(\bm Y;\bm\theta)\sqrt{C_2}~\big)}{T_{(\nu+np)}\big(\Delta(\bm Y;\bm\theta)\sqrt{C_0}~\big)}
\end{align}
where
\begin{align*}
 C_0=\frac{\nu+np}{\delta(\bm Y;\bm M,\bm\Sigma,\bm\Psi)+\nu-\Delta^2(\bm Y;\bm\theta)},\ \ \
 C_2=\frac{\nu+np+2}{\delta(\bm Y;\bm M,\bm\Sigma,\bm\Psi)+\nu-\Delta^2(\bm Y;\bm\theta)}.
\end{align*} 
Additionally, using the law of iterated expectations, we can obtain
\begin{align}
E(\gamma~W\vert\bm Y)=\frac{\eta(\bm Y;\bm\theta)}{\rho(\bm\Sigma,\bm\Psi,\bm\Lambda)+1}E(W\vert\bm Y)+\frac{1}{\sqrt{\rho(\bm\Sigma,\bm\Psi,\bm\Lambda)+1}}\zeta(\bm Y)
\end{align}
and
\begin{align}
E(\gamma^2~W\vert\bm Y)&=\frac{1}{\rho(\bm\Sigma,\bm\Psi,\bm\Lambda)+1}+\frac{\eta^2(\bm Y;\bm\theta)}{\big(\rho(\bm\Sigma,\bm\Psi,\bm\Lambda)+1\big)^2}E(W\vert\bm Y)\notag\\&+\frac{\eta(\bm Y;\bm\theta)}{\big(\rho(\bm\Sigma,\bm\Psi,\bm\Lambda)+1\big)^{3/2}}\zeta(\bm Y),
\end{align}
where 
\begin{align*}
\zeta(\bm Y)&=\frac{\Gamma\big(\frac{\nu+np+1}{2}\big)}{\sqrt{2\pi}~T_{(\nu+np)}\big(\Delta(\bm Y;\bm\theta)\sqrt{C_0}~\big)\Gamma\big(\frac{\nu+np}{2}\big)}\bigg(\frac{\delta(\bm Y;\bm M,\bm\Sigma,\bm\Psi)+\nu}{2}\bigg)^{-\frac{\nu+np+1}{2}}\\&\times\bigg(\frac{\delta(\bm Y;\bm M,\bm\Sigma,\bm\Psi)+\nu-\Delta^2(\bm Y;\bm\theta)}{2}\bigg)^{\frac{\nu+np}{2}}.
\end{align*}

\subsection{Parameter estimation via the ECME algorithm}\label{sub_EM}

Suppose $\bm{Y}=(\bm{Y}_1,\ldots,\bm{Y}_N)$ constitutes a set of $n\times p$-dimensional observed samples of size $N$ arising from the MVST model.
In the EM framework, the latent variables are $\bm{w}=(w_1,\ldots,w_N)$ and $\bm\gamma=(\gamma_1,\ldots,\gamma_N)$.
With these, the complete data is given by $\bm{Y}_c=(\bm{Y},\bm{w},\bm\gamma)$. 

According to  \eqref{joint_pdf}, the log-likelihood function of $\bm\theta$
corresponding to the complete-data $\bm{Y}_c$, excluding additive constants and terms that do not involve parameters of the model, is given by
\begin{align}\label{ell_MVST}
\ell_{c}(\bm{\theta}\mid \bm{Y}_c)&=\frac{1}{2}\sum_{i=1}^N\bigg\{\nu\log\bigg(\frac{\nu}{2}\bigg)-2\log\Gamma\bigg(\frac{\nu}{2}\bigg)-p\log\vert\bm\Sigma\vert-n\log\vert\Psi\vert\notag\\&+2\eta(\bm Y_i;\bm\theta)\gamma_iw_i-\bigg(\rho(\bm\Sigma,\bm\Psi,\bm\Lambda)+1\bigg)\gamma_i^2w_i\notag\\&-\bigg(\delta(\bm Y_i;\bm M,\bm\Sigma,\bm\Psi)+\nu\bigg)w_i+\big(\nu+np-1\big)\log w_i\bigg\}.
\end{align}

In the $k$th iteration, the E-step requires the calculation of the so-called $Q$-function, which is
the conditional expectation of \eqref{ell_MVST}, given the observed data $\bm Y$ and
the current estimate $\hat{\bm\theta}^{(k)}$, where the superscript $^{(k)}$ denote the updated estimates at $k$th step of the iterative process. To evaluate the $Q$-function,
we then need the following conditional expectations:
\begin{align}\label{e_step}
\hat{w}_{i}^{(k)}=E(W_i\mid \bm Y_i,\hat{\bm\theta}^{(k)}),\quad \hat{\kappa}_{1i}^{(k)}=E(\gamma_iW_i\mid \bm Y_i,\hat{\bm\theta}^{(k)}),\quad \hat{\kappa}_{2i}^{(k)}=E(\gamma_i^2W_i\mid \bm Y_i,\hat{\bm\theta}^{(k)}),
\end{align}
which have explicit expressions as given earlier and the term
\begin{equation}
 \hat{\kappa}_{3i}^{(k)}=E(\log W_i\mid \bm Y_i,\hat{\bm\theta}^{(k)}),\label{k3}
\end{equation}
which is difficult to evaluate explicitly. So, we perform the ECME algorithm to avoid computing the expectation in \eqref{k3}.

Substituting \eqref{e_step} and \eqref{k3} into \eqref{ell_MVST}, we obtain the following expression for the $Q$-function:
\begin{align}\label{q_MVST}
Q(\bm{\theta}\mid \hat{\bm{\theta}}^{(k)})&=\frac{1}{2}\sum_{i=1}^N\bigg\{\nu\log\bigg(\frac{\nu}{2}\bigg)-2\log\Gamma\bigg(\frac{\nu}{2}\bigg)-p\log\vert\bm\Sigma\vert-n\log\vert\Psi\vert\notag\\&+2\eta(\bm Y_i;\bm\theta)\hat{\kappa}_{1i}^{(k)}-\bigg(\rho(\bm\Sigma,\bm\Psi,\bm\Lambda)+1\bigg)\hat{\kappa}_{2i}^{(k)}\notag\\&-\bigg(\delta(\bm Y_i;\bm M,\bm\Sigma,\bm\Psi)+\nu\bigg)\hat{w}_i^{(k)}+\big(\nu+np-1\big)\hat{\kappa}_{3i}^{(k)}\bigg\}.
\end{align}

The CM-steps are implemented to update estimates of $\bm\theta$ in the order of $\bm M$, $\bm\Sigma$, $\bm\Psi$, $\bm\Lambda$ and $\nu$
by maximizing, one by one, the $Q$-function obtained in the E-step. After some algebraic manipulations, they are summarized in the following CMQ and CML steps:

\noindent{\bf CMQ-step 1:} Fixing $\bm\Lambda=\hat{\bm\Lambda}^{(k)}$, we update $\hat{\bm M}^{(k)}$  by maximizing \eqref{q_MVST} with respect to $\bm M$, leading to
\begin{equation*}
\hat{\bm M}^{(k+1)}=\frac{\sum_{i=1}^{N}\hat{w}_i^{(k)}\bm Y_i-\hat{\bm\Lambda}^{(k)}\sum_{i=1}^{N}\hat{\kappa}_{1i}^{(k)}}{\sum_{i=1}^{N}\hat{w}_i^{(k)}};
\end{equation*}

\noindent{\bf CMQ-step 2:} Fixing $\bm M=\hat{\bm M}^{(k+1)}$,$\bm\Psi=\hat{\bm\Psi}^{(k)}$ and $\bm\Lambda=\hat{\bm\Lambda}^{(k)}$, then update $\hat{\bm\Sigma}^{(k)}$ by maximizing \eqref{q_MVST} over $\bm\Sigma$, yielding
\begin{align*}
\hat{\bm\Sigma}^{(k+1)}&=\frac{1}{Np}\bigg\{\sum_{i=1}^N \hat{w}_i^{(k)}\big(\bm Y_i-\hat{\bm M}^{(k+1)}\big)\hat{\bm\Psi}^{-1(k)}\big(\bm Y_i-\hat{\bm M}^{(k+1)}\big)^\top\\&+\hat{\bm\Lambda}^{(k)}\hat{\bm\Psi}^{-1(k)}\hat{\bm\Lambda}^{\top(k)}\sum_{i=1}^N\hat{\kappa}_{2i}^{(k)}-\sum_{i=1}^N\hat{\kappa}_{1i}^{(k)}\big(\bm Y_i-\hat{\bm M}^{(k+1)}\big)\hat{\bm\Psi}^{-1(k)}\hat{\bm\Lambda}^{\top(k)}\\&-\hat{\bm\Lambda}^{(k)}\hat{\bm\Psi}^{-1(k)}\sum_{i=1}^N\hat{\kappa}_{1i}^{(k)}\big(\bm Y_i-\hat{\bm M}^{(k+1)}\big)^\top\bigg\};
\end{align*}

\noindent{\bf CMQ-step 3:} Fixing $\bm M=\hat{\bm M}^{(k+1)}$,$\bm\Sigma=\hat{\bm\Sigma}^{(k+1)}$ and $\bm\Lambda=\hat{\bm\Lambda}^{(k)}$, we update $\hat{\bm\Sigma}^{(k)}$ by maximizing \eqref{q_MVST} over $\bm\Psi$, yielding
\begin{align*}
\hat{\bm\Psi}^{(k+1)}&=\frac{1}{Nn}\bigg\{\sum_{i=1}^N \hat{w}_i^{(k)}\big(\bm Y_i-\hat{\bm M}^{(k+1)}\big)^\top\hat{\bm\Sigma}^{-1(k)}\big(\bm Y_i-\hat{\bm M}^{(k+1)}\big)\\&+\hat{\bm\Lambda}^{\top(k)}\hat{\bm\Sigma}^{-1(k)}\hat{\bm\Lambda}^{(k)}\sum_{i=1}^N\hat{\kappa}_{2i}^{(k)}-\sum_{i=1}^N\hat{\kappa}_{1i}^{(k)}\big(\bm Y_i-\hat{\bm M}^{(k+1)}\big)^\top\hat{\bm\Sigma}^{-1(k)}\hat{\bm\Lambda}^{(k)}\\&-\hat{\bm\Lambda}^{\top(k)}\hat{\bm\Sigma}^{-1(k)}\sum_{i=1}^N\hat{\kappa}_{1i}^{(k)}\big(\bm Y_i-\hat{\bm M}^{(k+1)}\big)\bigg\};
\end{align*}

\noindent{\bf CMQ-step 4:} Fixing $\bm M=\hat{\bm M}^{(k+1)}$, we obtain $\hat{\bm\Lambda}^{(k+1)}$ by maximizing (\ref{q_MVST}) over $\bm\Lambda$, yielding
\begin{align*}
\hat{\bm\Lambda}^{(k+1)}=\frac{\sum_{i=1}^{N}\hat{\kappa}_{1i}^{(k)}\big(\bm Y_i-\hat{\bm M}^{(k+1)}\big)}{\sum_{i=1}^{N}\hat{\kappa}_{2i}^{(k)}}.
\end{align*}

An update of $\hat{\nu}^{(k)}$ can be achieved
by directly maximizing the constrained actual log-likelihood function. This gives rise to the following CML-step:

\noindent{\bf CML-step:} Update $\hat{\nu}^{(k)}$ by optimizing the following constrained log-likelihood function:
\begin{equation*}
\hat{\nu}^{(k+1)}=\arg\max_{\nu}\sum_{i=1}^N\log f_{MVST}\big(\bm Y_i; \hat {\bm M}^{(k+1)} ,\hat{\bm\Sigma}^{(k+1)},\hat{\bm\Psi}^{(k+1)},\hat{\bm\Lambda}^{(k+1)},\nu\big).
\end{equation*}

\section{Fitting finite mixtures of MVST distributions}\label{sec-FM}
\subsection{The model}
We consider $N$ independent random variables $\bm Y_1,\cdots , \bm Y_N$ observed from a $G$-component mixture of MVST distributions, given by
\begin{align*}
f(\bm Y_i;\bm{\Theta})=\sum_{g=1}^G \pi_g f_{MVST} (\bm Y_i;\bm\theta_g),
\end{align*}
where $\sum_{i=1}^G\pi_g=1$ and $\bm{\Theta}$ is the set containing all the parameters of the considered mixture model. To pose this mixture model as an incomplete data problem, we introduce allocation variables $\bm{Z}_i=(Z_{i1},\cdots,Z_{iG})$, where a particular element $Z_{ig}$ is equal to $1$ if $Y_i$ belongs to group $g$ and is equal to zero otherwise. 
Observe that $\bm{Z_i}$ follows a
multinomial random vector with $1$ trial and cell probabilities $\pi_1,\cdots,\pi_G$, denoted by $\bm{Z}_i\sim\mathcal{M}(1;\pi_1,\cdots,\pi_G)$.
The hierarchical representation in \eqref{hier_rep},  originally designed for single distribution, can be extended to the mixture modeling framework, as follows:
\begin{align}\label{hier_mix}
\bm Y_i\vert(\gamma_i,w_i,Z_{ig}=1)&\sim N_{n\times p}(\bm M_g+\gamma_i\bm\Lambda_g,w_i^{-1}\bm\Sigma_g,\bm\Psi_g),\notag\\
\gamma_i\vert (w_i,Z_{ig}=1)&\sim TN(0,w_i^{-1})I_{(0,\infty)},\\
W_i\vert Z_{ig}=1&\sim\Gamma(\nu_g/2,\nu_g/2),\notag\\
Z_{ig}&\sim\mathcal{M}(1;\pi_1,\cdots,\pi_G).\notag
\end{align}

\sloppy {It then follows from the hierarchical structure in \eqref{hier_mix}, on the basis of the observed data $\bm{Y}=(\bm Y_1,\ldots,\bm Y_N)$ and latent data $\bm w=(w_1,\ldots,w_N)$, $\bm{\gamma}=(\gamma_1,\ldots,\gamma_N)$ and $\bm{Z}=(Z_{1},\cdots,Z_{N})$, excluding additive constants, the complete data log-likelihood function of 
$\bm{\Theta}$ based on the complete data $\bm{Y}_c=(\bm Y,\bm w,\bm\gamma,\bm Z)$ is}
\begin{align}\label{ell_mixMVST}
\ell_{c}(\bm{\theta}\mid \bm{Y}_c)&=\frac{1}{2}\sum_{i=1}^N\sum_{g=1}^G Z_{ig}\bigg\{\nu_g\log\bigg(\frac{\nu_g}{2}\bigg)-2\log\Gamma\bigg(\frac{\nu_g}{2}\bigg)-p\log\vert\bm\Sigma_g\vert-n\log\vert\Psi_g\vert\notag\\&+2\eta(\bm Y_i;\bm\theta_g)\gamma_iw_i-\bigg(\rho(\bm\Sigma_g,\bm\Psi_g,\bm\Lambda_g)+1\bigg)\gamma_i^2w_i\notag\\&-\bigg(\delta(\bm Y_i;\bm M_g,\bm\Sigma_g,\bm\Psi_g)+\nu_g\bigg)w_i+\big(\nu_g+np-1\big)\log w_i\bigg\}.
\end{align}

The expected value of (\ref{ell_mixMVST}) to start the E-step, given current parameter $\bm{\Theta}^{(k)}$,  requires some conditional expectations, including
\begin{align}\label{E_step_mix}
\hat{z}_{ig}^{(k)}&=E(Z_{ig}\vert\bm Y_i,\hat{\bm{\Theta}}^{(k)})=\frac{\hat{\pi}_g^{(k)} f_{MVST}(\bm y_i;\hat{\bm{\theta}}_g^{(k)})}{f(\bm y_i;\hat{\bm{\Theta}}^{(k)})},\notag\\
 \hat{w}_{ig}^{(k)}&=E(W_i\mid \bm Y_i,Z_{ig}=1,\hat{\bm\Theta}^{(k)})=E(W_i\mid \bm Y_i,\hat{\bm\theta}_g^{(k)}),\\
\hat{\kappa}_{1ig}^{(k)}&=E(\gamma_iW_i\mid \bm Y_i,Z_{ig}=1,\hat{\bm\Theta}^{(k)})=E(\gamma_iW_i\mid \bm Y_i,\hat{\bm\theta}_g^{(k)}),\notag\\
\hat{\kappa}_{2ig}^{(k)}&=E(\gamma_i^2W_i\mid \bm Y_i,Z_{ig}=1,\hat{\bm\Theta}^{(k)})=E(\gamma_i^2W_i\mid \bm Y_i,\hat{\bm\theta}_g^{(k)})\notag
\end{align}
and 
\begin{equation*}
\hat{\kappa}_{3ig}^{(k)}=E(\log W_i\mid \bm Y_i,Z_{ig}=1,\hat{\bm\Theta}^{(k)})=E(\log W_i\mid \bm Y_i,\hat{\bm\theta}_g^{(k)})
\end{equation*}
for which we utilize the CML step as mentioned earlier.
Consequently, the conditional expectation of the complete data log-likelihood is obtained as
\begin{align}\label{q_mixMVST}
Q(\bm{\theta}\mid \hat{\bm{\theta}}^{(k)})&=\frac{1}{2}\sum_{i=1}^N\sum_{g=1}^G\hat{z}_{ig}^{(k)}\bigg\{\nu_g\log\bigg(\frac{\nu_g}{2}\bigg)-2\log\Gamma\bigg(\frac{\nu_g}{2}\bigg)-p\log\vert\bm\Sigma_g\vert-n\log\vert\Psi_g\vert\notag\\&+2\eta(\bm Y_i;\bm\theta_g)\hat{\kappa}_{1ig}^{(k)}-\bigg(\rho(\bm\Sigma_g,\bm\Psi_g,\bm\Lambda_g)+1\bigg)\hat{\kappa}_{2ig}^{(k)}\notag\\&-\bigg(\delta(\bm Y_i;\bm M_g,\bm\Sigma_g,\bm\Psi_g)+\nu_g\bigg)\hat{w}_i^{(k)}+\big(\nu_g+np-1\big)\hat{\kappa}_{3ig}^{(k)}\bigg\}.
\end{align}

Thus, the implementation of the ECME algorithm
proceeds as follows:

\noindent{\bf E-step:} Given $\bm{\Theta }$ $=\hat{\bm{\Theta }}^{(k)}$, compute $\hat{z}_{ig}^{(k)}$, $\hat{w}_{ig}^{(k)}$, $\hat{\kappa}_{1ig}^{(k)}$ and $\hat{\kappa}_{2ig}^{(k)}$ given in \eqref{E_step_mix}, for $i=1,\ldots,N$ and $g=1,\ldots,G$;

\noindent{\bf CM-step 1:} Calculate 
\begin{equation*}
\hat{\pi}_g=\frac{1}{N}\sum_{i=1}^N \hat{z}_{ig}^{(k)};
\end{equation*}

\noindent{\bf CM-step 2:} Update $\hat{\bm M}_g^{(k)}$and obtain $\hat{\bm M}_g^{(k+1)}$ as 
\begin{eqnarray*}
\hat{\bm M}_g^{(k+1)}=\frac{\sum_{i=1}^{N}\hat{z}_{ig}^{(k)}\hat{w}_i^{(k)}\bm Y_i-\hat{\bm\Lambda_g}^{(k)}\sum_{i=1}^{N}\hat{z}_{ig}^{(k)}\hat{\kappa}_{1ig}^{(k)}}{\sum_{i=1}^{N}\hat{z}_{ig}^{(k)}\hat{w}_i^{(k)}};
\end{eqnarray*}

\noindent{\bf CM-step 3:} Update $\hat{\bm\Sigma}_g^{(k)}$ as
\begin{align*}
\hat{\bm\Sigma}_g^{(k+1)}&=\frac{1}{p\sum_{i=1}^N\hat{z}_{ig}^{(k)}}\bigg\{\sum_{i=1}^N \hat{z}_{ig}^{(k)}\hat{w}_i^{(k)}\big(\bm Y_i-\hat{\bm M}_g^{(k+1)}\big)\hat{\bm\Psi}_g^{-1(k)}\big(\bm Y_i-\hat{\bm M}_g^{(k+1)}\big)^\top\\&+\hat{\bm\Lambda}_g^{(k)}\hat{\bm\Psi}_g^{-1(k)}\hat{\bm\Lambda}_g^{\top(k)}\sum_{i=1}^N\hat{z}_{ig}^{(k)}\hat{\kappa}_{2ig}^{(k)}-\sum_{i=1}^N\hat{z}_{ig}^{(k)}\hat{\kappa}_{1ig}^{(k)}\big(\bm Y_i-\hat{\bm M}_g^{(k+1)}\big)\hat{\bm\Psi}_g^{-1(k)}\hat{\bm\Lambda}_g^{\top(k)}\\&-\hat{\bm\Lambda}_g^{(k)}\hat{\bm\Psi}_g^{-1(k)}\sum_{i=1}^N\hat{z}_{ig}^{(k)}\hat{\kappa}_{1ig}^{(k)}\big(\bm Y_i-\hat{\bm M}_g^{(k+1)}\big)^\top\bigg\};
\end{align*}

\noindent{\bf CM-step 4:} Update $\hat{\bm\Psi}_g^{(k)}$ as
\begin{align*}
\hat{\bm\Psi}_g^{(k+1)}&=\frac{1}{n\sum_{i=1}^N\hat{z}_{ig}^{(k)}}\bigg\{\sum_{i=1}^N\hat{z}_{ig}^{(k)} \hat{w}_i^{(k)}\big(\bm Y_i-\hat{\bm M}_g^{(k+1)}\big)^\top\hat{\bm\Sigma}_g^{-1(k)}\big(\bm Y_i-\hat{\bm M}_g^{(k+1)}\big)\\&+\hat{\bm\Lambda}_g^{\top(k)}\hat{\bm\Sigma}_g^{-1(k)}\hat{\bm\Lambda}_g^{(k)}\sum_{i=1}^N\hat{z}_{ig}^{(k)}\hat{\kappa}_{2ig}^{(k)}-\sum_{i=1}^N\hat{z}_{ig}^{(k)}\hat{\kappa}_{1ig}^{(k)}\big(\bm Y_i-\hat{\bm M}_g^{(k+1)}\big)^\top\hat{\bm\Sigma}_g^{-1(k)}\hat{\bm\Lambda}_g^{(k)}\\&-\hat{\bm\Lambda}_g^{\top(k)}\hat{\bm\Sigma}_g^{-1(k)}\sum_{i=1}^N\hat{z}_{ig}^{(k)}\hat{\kappa}_{1ig}^{(k)}\big(\bm Y_i-\hat{\bm M}_g^{(k+1)}\big)\bigg\};
\end{align*}

\noindent{\bf CM-step 5:}
Update $\hat{\bm\Lambda}_g^{(k+1)}$ as
\begin{align*}
\hat{\bm\Lambda}_g^{(k+1)}=\frac{\sum_{i=1}^{N}\hat{z}_{ig}^{(k)}\hat{\kappa}_{1ig}^{(k)}\big(\bm Y_i-\hat{\bm M}_g^{(k+1)}\big)}{\sum_{i=1}^{N}\hat{z}_{ig}^{(k)}\hat{\kappa}_{2ig}^{(k)}};
\end{align*}

\noindent{\bf CML-step:} Update $\hat{\bm\nu}^{(k)}=(\hat{\nu}_1,\ldots,\hat{\nu}_G)$, by optimizing the  constrained log-likelihood function, as
\begin{equation*}
\hat{\bm\nu}^{(k+1)}=\arg\max_{\bm\nu}\sum_{i=1}^N\log \bigg(\sum_{g=1}^G\hat{\pi}_g^{(k)}f_{MVST}\big(\bm Y_i; \hat {\bm M}_g^{(k+1)} ,\hat {\bm\Sigma}_g^{(k+1)},\hat {\bm\Psi}_g^{(k+1)},\hat{\bm\Lambda}_g^{(k+1)},\nu_g\big)\bigg).
\end{equation*}
\subsection{Initialization}
In order to speed up the convergence process, it is important to establish a set of reasonable starting values.
To start the ECME algorithm for fitting the FM-MVST model, an intuitive scheme to partition data to $G$ components $\{\bm Y_g^{(0)}\}_{g=1}^G$, is to create an initial partition of data $\{\mbox{Vec}(\bm{Y_i})\}_{i=1}^N$ using the $K$-means algorithm \citep{macqueen1967some,lloyd1982least}. This would yield validate estimate of $\hat{z}^{(0)}_{ig}$ which in turn yields to $\hat{\pi}^{(0)}_g=\sum_{i=1}^N\hat{z}^{(0)}_{ig}/N$. Then, we
compute the sample mean, covariance matrix of rows and covariance matrix of columns of $\bm{Y}^{(0)}_g$ as good initial estimates
for $\hat{\bm M}^{(0)}_g$, $\hat{\bm\Sigma}^{(0)}_g$ and $\hat{\bm\Psi}^{(0)}_g$ as follows:
\begin{align*}
\hat{\bm M}^{(0)}_g&=\frac{\sum_{i=1}^N\hat{z}^{(0)}_{ig} \bm Y_i}{\sum_{i=1}^N\hat{z}^{(0)}_{ig}}, \\
\hat{\bm\Sigma}^{(0)}_g&=\frac{\sum_{i=1}^N\sum_{j=1}^p\hat{z}^{(0)}_{ig} \big(\bm y_{ij}-\bm m_{gj}^{(0)}\big)\big(\bm y_{ij}-\bm m_{gj}^{(0)}\big)^\top}{p\sum_{i=1}^N\hat{z}^{(0)}_{ig}},\\
\hat{\bm\Psi}^{(0)}_g&=\frac{\sum_{i=1}^N\sum_{r=1}^n\hat{z}^{(0)}_{ig} \big(\bm y_{i.r}-\bm m_{g.r}^{(0)}\big)^\top\big(\bm y_{i.r}-\bm m_{g.r}^{(0)}\big)}{n\sum_{i=1}^N\hat{z}^{(0)}_{ig}},
\end{align*}
where $\bm m_{gj}^{(0)}$ and $\bm m_{g.r}^{(0)}$ denote the $j$-th column and $r$-th row of $\bm M_g^{(0)}$, respectively, and $\bm y_{ij}$ and $\bm y_{i.r}$ are the $j$-th column and $r$-th row of $\bm Y_i$, respectively.
The initial component skewness matrix, $\bm\Lambda_g^{(0)}$, are taken to the values randomly selected in the interval $(-1,1)$.
Finally, we initialize $\hat{\bm\nu}^{(0)}_g$ by taking it to be small as 5 or 10.

\subsection{Identifiability}
Model identifiability is key to securing unique and consistent estimates of model parameters. With regard to the mixtures of MVST distributions, the estimates of $\bm\Sigma_g$ and $\bm\Psi_g$ are only unique up to a strictly positive constant. To resolve this issue, a constraint needs to be placed, such as setting the trace of $\bm\Sigma_g$ equal to $n$ \citep{viroli2011finite} or fixing $\vert\bm\Sigma_g\vert=1$ \citep{sarkar2020parsimonious}. Herein, we set the first diagonal element of $\bm\Sigma_g$ as $1$ \citep{gallaugher2018finite}. This scaling procedure can be implemented at each iteration or at convergence, and either method has minimal impact on the final estimates and classifications achieved. To obtain the final parameter estimates, the resulting $\bm\Sigma_g$
is divided by the first diagonal element of $\bm\Sigma_g$, and then $\bm\Psi_g$ is multiplied by the first diagonal element of $\bm\Sigma_g$.
\section{Empirical study}\label{sec_sim}
\subsection{Finite-sample properties of ML estimators}
\label{sec_simfin}
\begin{table}[!t]
\footnotesize\tabcolsep=3.5pt
\begin{center}
\caption{Average RMSE based on 500 replications for the evaluation of ML estimates.} \label{Tab1}
\begin{tabular}{clcccccccccccccccccccccccccc}
\midrule
Scenario&$N$ &$\pi_1$&$\bm M_1$ &$\bm M_2$ &$\bm\Sigma_1$& $\bm\Sigma_2$ & $\bm\Psi_1$ & $\bm\Psi_2$& $\bm\Lambda_1$ & $\bm\Lambda_2$& $\nu_1$ & $\nu_2$\\\hline
&250&0.031& 0.169& 0.119& 0.058& 0.045& 0.128& 0.078& 0.174& 0.118& 1.638& 3.629\\
\Lower{{\rm I}}&500& 0.020& 0.129& 0.087& 0.044& 0.041& 0.093& 0.064& 0.126& 0.088& 1.215& 3.128\\
&1000&0.015& 0.091& 0.059& 0.032& 0.031& 0.066& 0.052& 0.088& 0.056& 0.803& 2.712\\
&2000&0.011& 0.062& 0.044& 0.027& 0.028& 0.051& 0.043 &0.062& 0.044& 0.702& 2.230\\\hline
&250&0.035& 0.260& 0.193& 1.348& 0.376& 0.785& 0.440& 0.257& 0.187& 2.544& 2.426\\
\Lower{{\rm II}}&500&0.022& 0.176& 0.130& 1.257& 0.338& 0.771& 0.425& 0.179& 0.129& 2.297& 2.219\\
&1000&0.014& 0.120& 0.097& 1.206& 0.312& 0.764& 0.423& 0.124& 0.097& 1.962& 1.897\\
&2000&0.011& 0.088& 0.068& 1.184& 0.302& 0.753& 0.401& 0.090 &0.070& 1.460& 1.388\\
\midrule
\end{tabular}
\end{center}
\end{table}

We conduct out here a simulation study for examining the accuracy of parameter estimates obtained by using the proposed ECME algorithm in Section~\ref{sub_EM}. We generate 500 Monte Carlo samples of sizes $n=250$, 500, 1000 and 2000 from the two component FM-MVST model, under two scenarios (low and moderate dimensional) described in Appendix~A.

The accuracy of the obtained parameter estimates are assessed by the average of the root mean squared errors (RMSE) of the elements of each estimated parameter.
The results shown in Table~\ref{Tab1} support a good performance of the proposed estimation method. Regardless of the considered scenario, it is seen that the RMSE values all tend to zero with increasing sample size, indicating satisfactory asymptotic properties of the ML estimates obtained by the proposed ECME algorithm. 

\subsection{Comparison of classification accuracy}
To examine the classification accuracy of the FM-MVST
 model, we generate 1000 samples from each of the
scenarios given in Appendix~A.
Under each scenario, we compared the FM-MVST model described in Section \ref{sec-FM} with the finite mixtures of matrix-variate normal (FM-MVN) and matrix-variate $t$ (FM-MVT) distributions, which are readily available in the R package {\tt MixMatrix}. Additionally, the finite mixtures of the reduced RMVSN (FM-RMVSN) distributions were fitted as a sub-model of FM-MVST model.

Model performance was assessed by comparing the classification accuracy and model selection criteria for all fitted models. For classification accuracy, we report the adjusted rand index (ARI) \citep{hubert1985comparing} which ranges
from 0 (no match) to 1 (perfect match) and the misclassification rate (MCR) of the MAP clustering for each model. Furthermore, the Bayesian information criterion (BIC) \citep{schwarz1978estimating} value is also reported as a model selection criteria.

We ran 100 simulations for each scenario and computed
the classification accuracy and model selection criteria for each simulation. Table \ref{tab2} presents the average BIC, ARI and MCR values along with their  standard errors. As one would expect, the model choice criteria selects the true model from which the data are generated. 

\section{Real data analysis}\label{sec_data}
In this section, we illustrate the proposed methodology by applying it to three well-known real datasets. 
\begin{table}[!t]
\begin{center}
\caption{Simulation results, based on 100 replications, for performance comparison of four mixture models under two scenarios. }\label{tab2}
\begin{tabular}{llccccccc}
\midrule
Scenario&Model& BIC&Std&ARI&Std& MCR&Std \\[2pt]
\hline
  & FM-MVN & 55419.78 & 831.20 & 0.82 & 0.17 & 0.05 & 0.05 \\ 
 \Lower{{\rm I}}  &FM-MVT & 45004.95 & 787.39 & 0.90 & 0.20 & 0.03& 0.06 \\ 
       & FM-RMVSN & 40103.90 & 962.33 & 0.95 & 0.18 & 0.02 & 0.05 \\  
       &FM-MVST & 38170.98 & 804.07 & 0.98 & 0.05 & 0.01 & 0.01
\\\hline
  & FM-MVN & 85450.81 & 705.80 & 0.91 & 0.15 & 0.07 & 0.06 \\ 
 \Lower{{\rm II}}  &FM-MVT & 76011.44 & 720.48 & 0.93 & 0.14 & 0.08& 0.05 \\ 
       & FM-RMVSN & 72917.04 & 703.48 & 0.94 & 0.12 & 0.05 & 0.02 \\  
       &FM-MVST & 69839.90 & 673.31 & 0.97 & 0.09 & 0.02 & 0.01 \\\midrule
\end{tabular}
\end{center}
\end{table}
\subsection{Landsat data}
The first application concerns the Landsat data (LSD), originally obtained by NASA, and available at Irvine machine learning repository (\url{http://archive.ics.uci.edu/ml}). Multi-spectral satellite imagery allows for multiple observations over a spatial grid, resulting in matrix-valued observations.  The LSD comprises of lines that consist of four spectral values representing nine pixel neighborhoods in a satellite image. Essentially, each line corresponds to a $4\times 9$ observation matrix. Additionally, every observation matrix in the LSD is classified into one of six distinct categories: red soil, cotton crop, grey soil, damp grey soil, soil with vegetation stubble, and very damp grey soil. For our analysis, we concentrate on three specific categories, red soil, grey soil, and soil with vegetation stubble, which have sizes of 461, 397 and 237, respectively.

Table~\ref{tab3} presents a summary of ML fitting results, including the maximized log-likelihood values, BIC, ARI and MCR of the four fitted models. The results reveal that the log-likelihood value for the FM-MVN distribution is lower than those for the FM-MVT distribution, indicating a poorer fit. In contrast, the skewed distributions (FM-MVST and FM-RMVSN) outperform their respective models. Particularly noteworthy is the superior performance of the FM-MVST model. The estimated tailedness parameters are $\nu_1=0.47$, $\nu_2=0.44$ and $\nu_3=0.58$, indicating a distribution of matrix observations characterized by a long-tailed behavior. 
\begin{SCtable}[][!t]
\caption{Summary results from fitting various models to the LSD data.}\label{tab3}
\centering
\small
\begin{tabular}{lccccc}
  \midrule
Model&G& Log-Likelihood& BIC& ARI&MCR \\[2pt]
\hline
FM-MVN&& --114954.90&   231799.40 & 0.67&0.14\\
FM-MVT&\Lower{3} & --113169.30& 228228.10 &0.69&0.13\\
FM-RMVSN&&--111213.50 & 225107.40& 0.76&0.09\\
FM-MVST&& --110836.60&  224374.60& 0.82&0.06\\
\midrule
\end{tabular}
\end{SCtable}
\subsection{Apes data}
The second application considers {\it apes} dataset included in the {\it shapes} R package \citep{Dryden-shapes}. The description of the dataset, is taken from \cite{dryden2016statistical}, is as follows. In an investigation to assess the cranial differences between the sexes of apes, 29 male
and 30 female adult gorillas (Gorilla), 28 male and 26 female adult chimpanzees
(Pan), and 30 male and 24 female adult orangutans (Pongo) were studied. Eight landmarks were chosen in the midline plane of each skull. These landmarks are anatomical landmarks and are located by an expert biologist. The dataset is stored as a list with two components: an array of coordinates
in eight landmarks in two dimensions for each skull ($8\times 2$ observation matrix and $N=167$), and a vector of group labels ($G=6$).

All the competing models are fitted for $G =6$ and their fitting results are reported in Table~\ref{tab4}.  It is clear that the FM-MNV model provide the worst fitting performance, whereas FM-MVST is the best model.
Similarly to the analyses of the previous Section, this may be an indication that the components of FM-MVN are not skewed and heavy tailed enough to adequately model the data. On a related note, the estimated tailedness parameters are $\nu_1=0.45$, $\nu_2=2.20$, $\nu_3=1.61$, $\nu_4=2.80$, $\nu_5=0.52$ and $\nu_6=1.61$, highlighting the presence of clusters with high levels of tailed behavior.
\begin{SCtable}[][!t]
\caption{Summary results from fitting various models to the Apes data.}\label{tab4}
\centering
\small
\begin{tabular}{lccccc}
  \midrule
Model&G& Log-Likelihood& BIC& ARI&MCR \\[2pt]
\hline
FM-MVN&& --7773.14&  17204.51 & 0.51&0.41\\
FM-MVT&\Lower{6} & --7609.66& 16877.54 &0.56&0.32\\
FM-RMVSN&&--6158.09 & 14522.03& 0.60&0.28\\
FM-MVST&& --5970.42&  14177.41& 0.67&0.25\\
\midrule
\end{tabular}
\end{SCtable}
\subsection{Melanoma data}
The performance of the FM-MVST model in skin cancer detection is demonstrated in the third and final application. The objective of the skin cancer detection project is to develop a framework for analyzing and assessing the risk of melanoma using dermatological photographs taken with a standard consumer-grade camera. Segmentation of the lesion is a crucial step for developing a skin cancer detection framework. The objective then is to find the border of the skin lesion. It is important that this step is performed accurately because many features used to assess the risk of melanoma are derived based on the lesion border. The set of images includes images extracted from the public databases DermIS and DermQuest, along with manual segmentations (ground truth) of the lesions is available at \url{https://uwaterloo.ca/vision-image-processing-lab/research-demos/skin-cancer-detection}.
\begin{figure}[!t]
\begin{center}
\includegraphics[height=8.5cm,width=12cm]{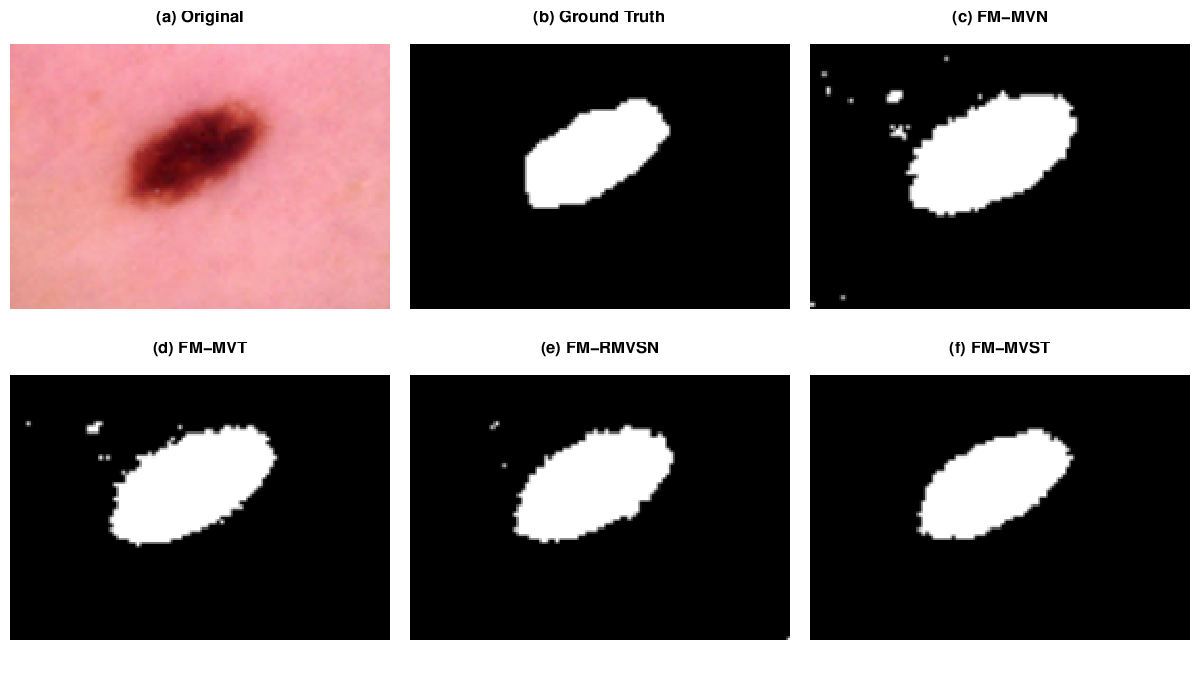}
\caption{Segmentation of lesion: (a) Original, (b) ground truth, and (c)-(f) segmented images obtained using different models. } \label{fig1}
\end{center}
\end{figure}

One skin image in $100\times 70$ pixels is displayed in Figure \ref{fig1} (a).
The objective now is to segment the image into two labels. Let us consider all pixels of three numerical RGB components denoting the red, green and blue intensities and a grayscale intensity, such as $y_i\in [0,255]^4$, which can be transformed into $[0,1]^4$. Upon considering each pixel as a $2\times 2$ matrix,  each pixel will then be grouped into $G=2$ clusters, where every cluster will be assumed to have a different distribution.

It follows from Table~\ref{tab5}, the FM-MVST model provides the best fit in terms of BIC as well as the lowest misclassification error for binary classification of each pixel. The estimates of tailedness parameters are $\nu_1=1.96$ and $\nu_2=0.92$, signifying the appropriateness
of the use of heavy-tailed $t$ distributions. Furthermore, the superiority of FM-MVST model is reflected visually in Figures \ref{fig1} (c)-(f), which depict the comparative segmentation performance of the fitted models in grayscale. 
The figures illustrate differences in identifying the lesion area and indicate that the proposed model exhibits a clearer boundary and a more consistent region of the lesion.
\begin{SCtable}[][!t]
\caption{Summary results from fitting various models to the Melanoma data.}\label{tab5}
\centering
\small
\begin{tabular}{lccccc}
  \midrule
Model&G& Log-Likelihood& BIC& ARI&MCR \\[2pt]\hline
FM-MVN&& 63941.64&  --127723.90 & 0.76&0.14\\
FM-MVT&\Lower{2} & 65509.39&--130859.40 &0.82&0.13\\
FM-RMVSN&&65601.15 & --130945.50& 0.91&0.12\\
FM-MVST&& 65630.64&  --130986.80& 0.95&0.09\\
\midrule
\end{tabular}
\end{SCtable}
 
\section{Concluding remarks}\label{sec:con}
We have introduced here a new family of matrix-variate distributions that can capture both skewness and heavy-tailedness simultaneously. This MVST model is based on a stochastic representation, which facilitates us to develop an efficient ECME algorithm for the maximum likelihood estimation of model parameters. We have evaluated the effectiveness and efficiency of the proposed method through two simulation studies. Additionally, we have used the proposed approach to analyze three real datasets, demonstrating its  capability in modeling asymmetric matrix-variate data. 
Future developments of this approach could include accommodating censored data and also in including different distributions for the variables $W$ and $U$ in the stochastic representation. Another interesting extension could involve incorporating the FM-MVST distribution into a mixture of regression framework. We have currently looking into these problems and hope to report the finding in a future paper.
\setcounter{equation}{0}
\section*{Appendix A}\label{APP_A}
The parameters used to generate data in Section~\ref{sec_simfin} are given in the following Table. Here, $\bm 1_p$ is used to denote the vector of length $p$ with all its entries as $1$ and $\bm I_p$ to denote the p-dimensional identity matrix.
\begin{table}[!h]
\footnotesize\tabcolsep=4pt
\begin{center}
\caption{The parameters used in the generation of data (scenarios {\rm I} and {\rm II}).} \label{Tab6}
\begin{tabular}{ccll}
\midrule
Scenario &Parameter &  Component 1 & Component 2\\\hline
&$\pi_g$ & 0.3 & 0.7\\
&$\bm M_g$ &$\left[ \begin{array}{cccc}
 -1  & \ \ 1 & -1&  2\\
 0 & \ \ 2 &  -1  &  0 \\
 0 & \ \  0 &   0  &-1
  \end{array}\right]$ &  $\left[ \begin{array}{cccc}
 0  & \ \ 2  & \ \ 0  & \  1\\
 0  &  \ 2  & \ 0  & -1 \\
 0  & \ 1  & \ 1   &-1
 \end{array}\right]$  \\
\Lower{\Lower{\Lower{${\rm I}$}}} &$\bm\Sigma_g$ & $\left[ \begin{array}{ccc}
 1.0 &0.0 &0.0\\
 0.0 & 0.7& -0.1 \\
 0.0 &-0.1&  1.0
 \end{array}\right]$ & $\left[ \begin{array}{ccc}
 1.0 & 0.1&  0.2\\
 0.1 & 0.5 &-0.5 \\
 0.2 &-0.5  &1.4
 \end{array}\right]$ \\
& $\bm\Psi_g$ & $\left[ \begin{array}{cccc}
 0.7  &0.0&  0.0 & 0.0\\
 0.0 & 1.0& -0.5 & 0.5 \\
 0.0 &-0.5&  1.5&  0.1\\
 0.0 & 0.5 & 0.1 & 1.0
 \end{array}\right]$ & $\left[ \begin{array}{cccc}
 1.0 & \  0.5&  \  0.0 \  & 0.0\\
 0.5 &  \ 1.0 &  \ 0.5 &  \ 0.5 \\
 0.0 &  \ 0.5 & \  1.0 & \  0.1\\
 0.0 &  \ 0.5& \   0.1 & \  1.0
 \end{array}\right]$\\
& $\bm\Lambda_g$ & $\left[ \begin{array}{cccc}
 1  &  \  -2& 0 &  \ 1\\
 1  &  \  -2 &0&  \ 1 \\
 1  &  \  -2 &  \ 0 & \ 1
 \end{array}\right]$ & $\left[ \begin{array}{cccc}
 0  &  \  1& -1 &  \ 0\\
 0  &  \  1 &-1&  \ -1 \\
 1  &  \  1 &  \ 0 & \ -1
 \end{array}\right]$ \\
& $\nu_g$ & 3 & 5\\\hline
&$\pi_g$ & 0.4 & 0.6\\
&$\bm M_g$ &$\left[ \begin{array}{cc}
  -1 & -1\\
  0 & \ 1  
  \end{array}\right]$ $\otimes\bm 1_5$ &  $\left[ \begin{array}{cc}
  0  & \ \  0  \\
  2  &  \ \ 1  
 \end{array}\right]$ $\otimes\bm 1_5$   \\
\Lower{\Lower{\Lower{${\rm II}$}}} &$\bm\Sigma_g$ & $\left[ \begin{array}{cc}
 5.0 &-0.5\\
 -0.5 & \ 1.0 
 \end{array}\right]$ $\otimes\bm I_5$ & $\left[ \begin{array}{cc}
 2.0 & \ \ 0.1\\
 0.1 &\ \  0.5 
 \end{array}\right]$ $\otimes\bm{I}_5$ \\
& $\bm\Psi_g$ & $\left[ \begin{array}{cc}
  0.5 &\  0.0\\
 0.0 & \  0.5 
 \end{array}\right]$ & $\left[ \begin{array}{cc}
 1.0 & \  0.5  \\
 0.5 &  \ 1.0  
 \end{array}\right]$ \\
& $\bm\Lambda_g$ & $\left[ \begin{array}{cc}
  -2 &   \ \ 1\\
  -2  &  \  \ 1
 \end{array}\right]$ $\otimes\bm 1_5$ & $\left[ \begin{array}{cc}
 1  &\ \ 2 \\
 1  & \ \ 2
 \end{array}\right]$ $\otimes\bm 1_5$\\
& $\nu_g$ & 4 & 4\\
  \midrule
\end{tabular}
\end{center}
\end{table}
\clearpage
\bibliographystyle{plainnat}
\bibliography{Reference}
\end{document}